\documentclass[jap,reprint,
amsmath,amssymb,letterpaper,%
superscriptaddress,
 balancelastpage,raggedbottom,%
 citeautoscript,floatfix]
{revtex4-1}

\usepackage[usenames,dvipsnames,table]{xcolor}
\usepackage[
  breaklinks,             %
  bookmarks = false, %
  pdfpagemode   = UseNone,%
  pdfstartview 	= FitH,		%
  pdfstartpage  = 1,      %
  colorlinks    = true   %
]{hyperref}
\hypersetup{
    linkcolor=RubineRed,          
    citecolor=Blue,        
    filecolor=Mulberry,      
    urlcolor=RoyalBlue           
}
\usepackage{microtype}
\usepackage{graphicx}
\usepackage{natbib}
\usepackage{multirow}
\usepackage{enumerate}
\usepackage{setspace} 
\renewcommand{\deg}{\ensuremath{^{\circ}}}
\newcommand{\ex}{$\times ~10^{-6}$ K$^{-1}$}

\begin{document}

\title{High-temperature neutron diffraction and first-principles
study of temperature-dependent crystal structures and 
atomic vibrations in Ti$_3$AlC$_2$, Ti$_2$AlC, and  
Ti$_5$Al$_2$C$_3$}
\author{Nina J.\ Lane}
	\affiliation{Department\! of\! Materials\! Science\! \&\! Engineering,\! 
Drexel\! University,\! Philadelphia,\! PA 19104,\! USA}%
\author{Sven C.\ Vogel}
	\affiliation{Los\! Alamos\! Neutron\! Science\! Center,\! 
Los\! Alamos\! National\! Laboratory,\! Los\! Alamos,\! NM\! 87545,\! USA}%
\author{El'ad N.\ Caspi}
\thanks{On sabbatical leave from Nuclear Research Centre - Negev, Beer-Sheva, Israel}
	\affiliation{Department\! of\! Materials\! Science\! \&\! Engineering,\! 
Drexel\! University,\! Philadelphia,\! PA 19104,\! USA}%
\author{Michel W.\ Barsoum}%
\email[Michel W. Barsoum: ]{barsoumw@drexel.edu}
	\affiliation{Department\! of\! Materials\! Science\! \&\! Engineering,\! 
Drexel\! University,\! Philadelphia,\! PA 19104,\! USA}%

\date{\today}

\begin{abstract}

Herein we report on the thermal expansions and temperature-dependent 
crystal structures of select ternary carbide MAX phases 
in the Ti--Al--C phase diagram in the 100--1000$\deg$C temperature 
range. A bulk sample containing $38(\pm1)$ wt.\% Ti$_5$Al$_2$C$_3$ 
(``523''), $32(\pm1)$ wt.\%  Ti$_2$AlC (``211''), $18(\pm1)$ wt.\% 
Ti$_3$AlC$_2$ (``312''), and $12(\pm1)$ wt.\% 
(Ti$_{0.5}$Al$_{0.5}$)Al is studied by Rietveld analysis of high-temperature 
neutron diffraction data. We also report on the same for a single-phase sample of Ti$_3$AlC$_2$ for comparison. 
The thermal expansions of all the MAX phases studied are 
higher in the $c$ direction than in the $a$ direction. The bulk 
expansion 
coefficients --  9.3($\pm 0.1$) \ex~ for Ti$_5$Al$_2$C$_3$, 9.2($\pm 0.1$) \ex~ for 
Ti$_2$AlC, and 9.0($\pm 0.1$) \ex~ for Ti$_3$AlC$_2$ -- are comparable 
within one standard deviation of each other. In Ti$_5$Al$_2$C$_3$, the 
dimensions of the Ti--C octahedra for the 211-like and 312-like regions 
are comparable to the Ti--C octahedra in Ti$_2$AlC and 
Ti$_3$AlC$_2$, respectively.
The isotropic mean-squared atomic displacement parameters
are highest for the Al atoms in all three phases, and the values predicted
from first-principles phonon calculations agree well with those measured.  

\end{abstract}

\maketitle

\section{Introduction}

\begin{figure}
\centering
\includegraphics[width=\columnwidth,clip]{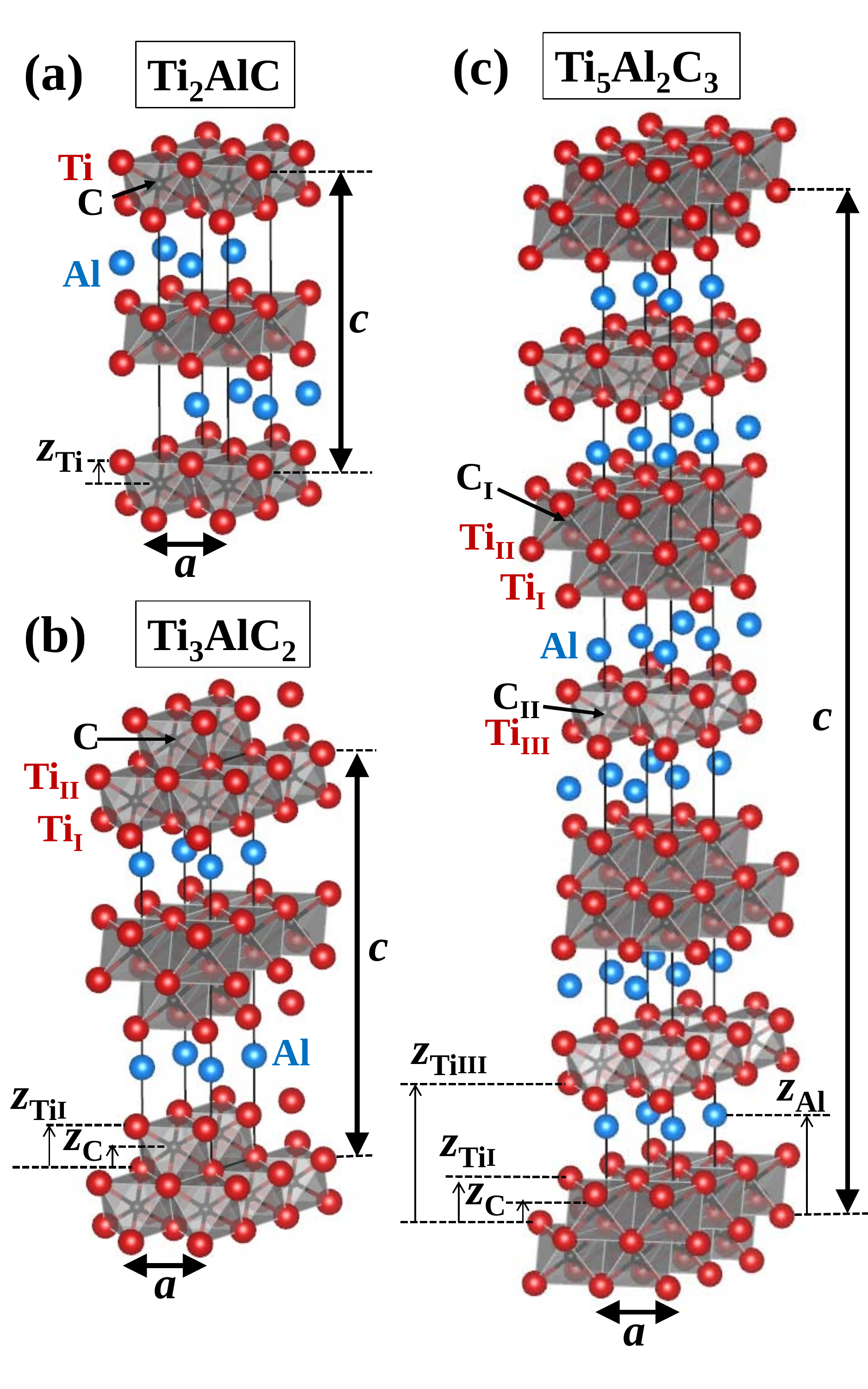}\vspace{-1em}
\caption{
Structure of (a) Ti$_2$AlC, (b) Ti$_3$AlC$_2$, and (c) Ti$_5$Al$_2$C$_3$ showing lattice parameters $a$ and $c$ and the unique Ti, Al, and C lattice sites as constrained by symmetry, along with the refined atomic position $z$ parameters.
\vspace{-1em}}
\label{fig:523struct}
\end{figure}

Titanium carbide is one the hardest transition metal binary carbides 
known, which renders it resistant to wear and makes it a favorable material for 
applications such as drill bits and cutting tools. With a melting 
point of over 3000$\deg$C, it is thermally quite stable. However, 
transition metal binary carbides are brittle, difficult to machine, and 
highly susceptible to thermal shock. The ternary carbides known as 
$M_{n+1}AX_n$ (MAX) phases \cite{BarsoumMAX2000} (where $n$ = 1,2, or 3 and $M$ is a 
transition metal, $A$ is a group A element mostly from groups 13 and 
14, and $X$ is either C or N) often overcome some of these 
shortcomings, while still being heat tolerant. As a class, the MAX phases 
have unusual -- yet attractive and often unique -- combinations of 
properties that bring together some of the best attributes of ceramics and 
metals \cite{Barsoum1996,ElRaghy1997,
BarsoumProgRep1997}.
Like metals, they are excellent electric and thermal 
conductors, with exceptional thermal shock resistance and damage tolerance \cite{Barsoum1996,ElRaghy1997,ganguly}.
In some cases, they are creep \cite{darincreep,RadovicCreep2001,Barsoum2003Creep,compressivecreep}, oxidation \cite{BarsoumOxidation1997,Sundberg}, and fatigue \cite{lifatigue} resistant.  
Furthermore, they are elastically quite stiff, yet readily machinable \cite{AmSci}.   
In this work we are interested in the MAX phases in the Ti--Al--C 
system. Both Ti$_2$AlC [Fig. \ref{fig:523struct}(a)] and Ti$_3$AlC$_2$ 
[Fig. \ref{fig:523struct}(b)] have been relatively well 
studied. The crystal structure of Ti$_2$AlC 
was first solved in the 1960s \cite{JeitschkoTi2AlC};
Ti$_3$AlC$_2$ was discovered several decades later in 1994 \cite{PietzkaSchuster}.
Of the $>$ 60 
MAX phases known to date, Ti$_2$AlC and 
Ti$_3$AlC$_2$ are particularly attractive in terms of high temperature applications.
They are two of the most lightweight and oxidation 
resistant MAX phases \cite{CorrosionScience,Wangoxidation}, and the accessibility and relative low cost of their raw materials 
render them the most promising for up-scaling and industrialization.

In addition to Ti$_2$AlC, henceforth referred to as 211, and 
Ti$_3$AlC$_2$, henceforth referred to as 312, there is also a 
Ti$_5$Al$_2$C$_3$ -- or 523 -- phase, which is in a category of 
higher-order MAX phases.  The latter were first
reported in 2004 in the Ti--Si--C system (\textit{i.e.} Ti$_5$Si$_2$C$_3$ and 
Ti$_7$Si$_2$C$_5$) by Palmquist et al \cite{Palmquist2004}.
Wilhelmsson et al. \cite{Wilhelmsson} reported the existence of
Ti$_5$Al$_2$C$_3$ in 2006.  In both cases, these phases were only observed
in local 
stacking sequences  in transmission electron microscopy, TEM, micrographs. 

More recently, the stacking sequence of Ti$_5$Al$_2$C$_3$ was 
characterized using X-ray diffraction, XRD, in two 
studies \cite{Wang523,523paper}. In our 
study \cite{523paper}, the characterized sample contained 
$43(\pm 2)$ wt.\% 
Ti$_5$Al$_2$C$_3$. In another study \cite{Wang523},
only a small 
amount of Ti$_5$Al$_2$C$_3$ was observed and 
neither weight nor volume fractions were reported. 
We note in passing that the first structure proposed by Wang \textit{et al.} in Ref. \cite{Wang523} is  totally wrong and unsubstantiated by the results shown in that paper (see Ref. \cite{523comment}). It is also crucial to note that our paper was submitted a few weeks before that of Ref. \cite{Wang523}. Interestingly, Ref. \cite{Wang523} was submitted, reviewed, and accepted in exactly one week and published soon thereafter. It was only after our paper was published that the same group, working with a composition that was only 19.7 wt\% Ti$_5$Al$_2$C$_3$, made the case that the space group was $R\bar{3}m$ \cite{Zhang523}.

Like most other MAX phases, Ti$_2$AlC and Ti$_3$AlC$_2$ both have 
layered hexagonal structures [Figs. \ref{fig:523struct}(a) and \ref{fig:523struct}(b), respectively] belonging to 
space group $P6_3/mmc$ (No. 194). In both structures, Ti--C layers 
(Ti$_2$C for Ti$_2$AlC and Ti$_3$C$_2$ for Ti$_3$AlC$_2$) are 
interleaved between layers of Al. The Ti$_5$Al$_2$C$_3$ phase, on the 
other hand [Fig. \ref{fig:523struct}(c)], consists of alternating Ti$_2$C (``211-like'') 
and Ti$_3$C$_2$ (``312-like'') layers interleaved between Al layers. 
Due to the shift in stacking sequence, three formula units need to be included in a unit cell.
The lattice parameters are thus $a$=$3.064(\pm 0.002)$ \AA\, and $c$=$48.23(\pm 0.02)$ \AA~ \cite{523paper}.
While XRD and TEM have been used to characterize the room 
temperature lattice parameters of Ti$_5$Al$_2$C$_3$ and verify its 
stoichiometry and stacking sequence \cite{523paper},
 a complete structure 
refinement to determine bond lengths has not yet been performed.
Furthermore, its thermal expansion coefficients, TECs, have not been measured. Previously, high-temperature neutron diffraction, HTND, was used to investigate the high-temperature 
stability of Ti$_2$AlC \cite{thermalstability,Pang211}
 and Ti$_3$AlC$_2$ \cite{thermalstability,Pang312}
 up to 
1550$\deg$C, but in Refs. \cite{thermalstability} and \cite{Pang211}, only the temperature-dependent phase fractions were 
studied with no reports of high-temperature crystallographic data.  In \cite{Pang312}, the lattice parameters and TECs were reported for Ti$_3$AlC$_2$, but the bond lengths have yet to be reported for both Ti$_2$AlC and Ti$_3$AlC$_2$.


The original aim of this work was to determine the crystal structure parameters of phase-pure 
Ti$_2$AlC. However, for reasons that are not fully understood, during processing, Ti$_2$AlC 
decomposed to yield Ti$_5$Al$_2$C$_3$ and some Ti$_3$AlC$_2$ \cite{523paper}. 
And while obtaining the relevant crystal structure parameters on such a multiphase sample is not 
ideal we decided to proceed with this HTND study nevertheless for the following reasons: 
\begin{enumerate}[(a)]
\item  To date, neither we nor others have been able to synthesize phase-pure Ti$_5$Al$_2$C$_3$, 
and the results obtained here are 
better than no results; 
\item Given the very similar crystal structures,  
elastic properties, and TECs among the three phases (see 
below), the deviations from phase-pure behavior due to the presence of other phases 
should be small; 
\item Careful Rietveld analysis 
can deconovolute the contributions from each phase. 
\end{enumerate}
These comments notwithstanding, to 
estimate the error in the various parameters as a function of temperature, we compare the results 
obtained on the multiphase sample with a predominantly single-phase sample 
of Ti$_3$AlC$_2$ and show that the differences, for the most part, are equal to (or less than) the 
experimental uncertainty.

Herein we use HTND on a sample consisting of 
38($\pm$1) wt.\% Ti$_5$Al$_2$C$_3$ (``523''), $32 (\pm 1)$ wt.\% 
Ti$_2$AlC (``211''), $18 (\pm 1)$ wt.\% Ti$_3$AlC$_2$ (``312''), and 
$12(\pm 1)$ wt.\% of an additional intermetallic TiAl phase to 
determine the temperature-dependent crystal structures of all three 
Ti--Al--C carbide phases. 
We report on the temperature evolution of the lattice parameters, 
isotropic thermal atomic displacement parameters, ADPs, and bond lengths 
during both heating and cooling for 
Ti$_2$AlC and Ti$_3$AlC$_2$.  We verify our results 
by comparing the room temperature lattice parameters and lattice expansions to previous 
studies of predominantly single-phase Ti$_2$AlC and Ti$_3$AlC$_2$ 
samples \cite{PietzkaSchuster,Ti2AlN-CXray,2000Ti2AlN-C,ScabaroziCTE,Tzenov}.
We also present HTND results for a predominantly single-phase Ti$_3$AlC$_2$ 
sample for comparison.

\section{Experimental details}

\begin{figure*}
\centering
\includegraphics[width=0.90 \textwidth]{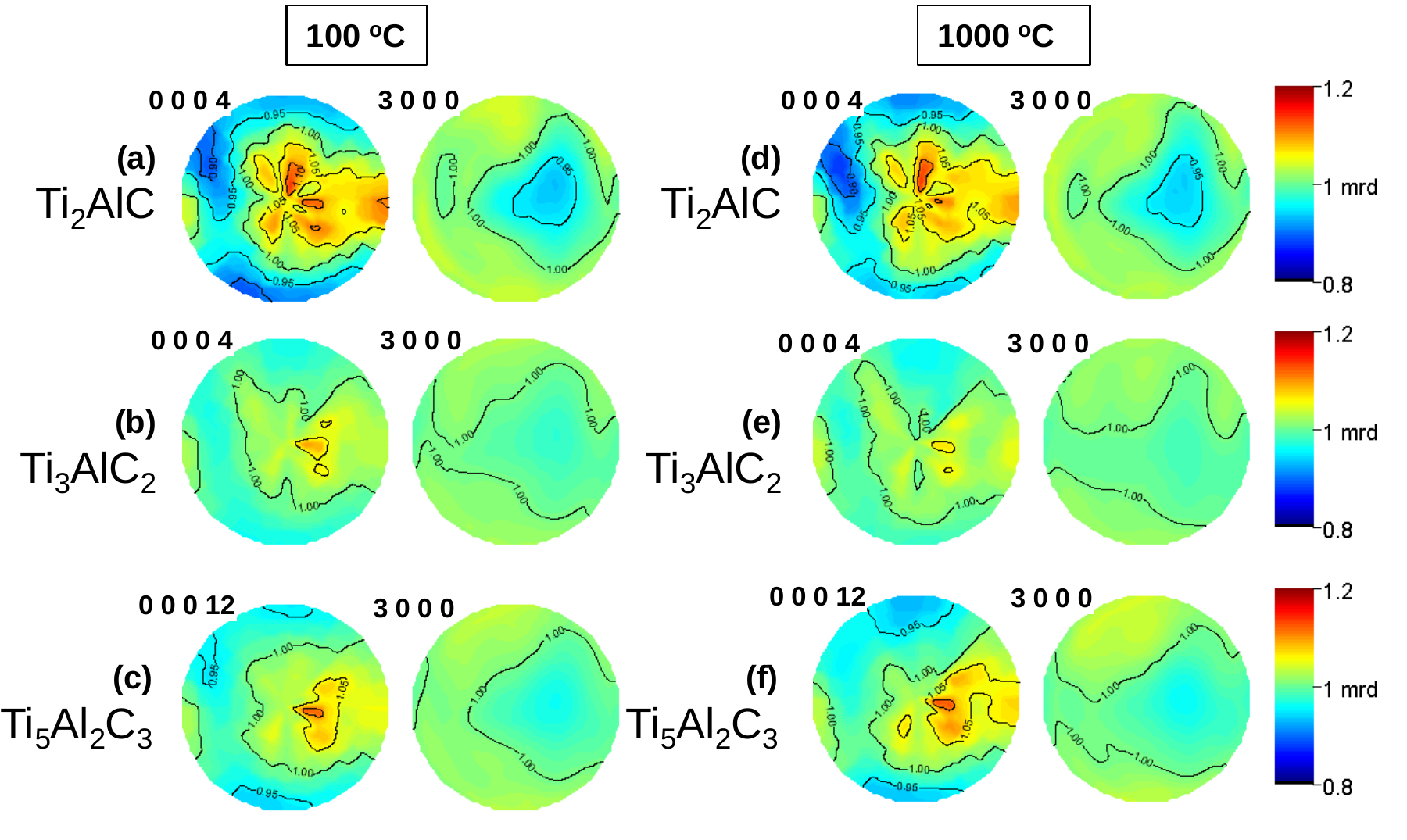}
\caption{Pole figure for 000$l$ and 3000 recalculated from the orientation distribution of HIPPO data for (a) Ti$_2$AlC, (b) Ti$_3$AlC$_2$, and (c) Ti$_5$Al$_2$C$_3$ at 100$\deg$C, and (d) Ti$_2$AlC, (e) Ti$_3$AlC$_2$, and (f) Ti$_5$Al$_2$C$_3$ at 1000$\deg$C. Sample cylinder axis is in the center of pole figures. }
\label{fig:523texture}
\end{figure*}

\subsection{Sample synthesis}

The sample used herein was prepared using 
pre-reacted Ti$_2$AlC
powders that were commercially obtained 
(Kanthal, Hallstahammar, Sweden). The powders were cold isostatic
pressed, CIPed, at 200MPa and heated at a rate of 300$\deg$C h$^{-1}$   to 1500$\deg$C,
 then sintered for 2 h under a hydrogen atmosphere.

The Ti$_3$AlC$_2$ sample was prepared by hot pressing, HPing, pre-reacted Ti$_2$AlC (Kanthal, Hallstahammar, Sweden) and titanium carbide (Alfa Aesar,Ward Hill, MA) in a 1:1 ratio to make 
3:1:2 stoichiometry of Ti:Al:C.  Powders were ball milled for 24 h, placed in a graphite die,
and heated in a graphite-heated hot press under a vacuum of 10$^{-1}$ Torr at a rate of 
500$\deg$C h$^{-1}$ to 1400$\deg$C.  It was held for 4 h under a pressure of $\sim 40$ MPa before cooling.

In both cases, bulk samples 9 mm in diameter and 3 cm high were used for the HTND experiments.


\subsection{High-temperature neutron diffraction}

The HTND experiments were conducted on the 
High-Pressure Preferred Orientation (HIPPO) neutron diffractometer 
\cite{HIPPO,HIPPO2,reiche2012furnace} at the 
Lujan Neutron Scattering Center, Los Alamos National Laboratory. 
For both the multiphase Ti--Al--C sample and Ti$_3$AlC$_2$, bulk 
samples were placed in a vanadium, V, holder (9 mm diameter, 
0.15 mm wall thickness), mounted in an ILL-type high-temperature vacuum 
furnace with a V setup, and heated at a rate of 20$\deg$C min$^{-1}$. Data were 
collected every 100$\deg$C during heating from 100$\deg$C to 1000$\deg$C.
For the multiphase sample only, data were collected every 
200$\deg$C upon cooling as well.  At each measurement, the temperature was 
held constant during data collection and neutrons were detected with 42 panels of $^3$He detector tubes arranged on five rings with nominal diffraction angles of 
39$\deg$, 60$\deg$, 90$\deg$, 120$\deg$, and 144$\deg$. To allow texture 
analysis, the sample was measured at rotation angles of 0$\deg$, 22$\deg$, 
and 45$\deg$ around 
the vertical axis for each temperature, with a count time of 15 min per orientation 
for a total count time of 45 min per temperature.

\begin{figure*}[t]
\centering
		\includegraphics[width= \textwidth]{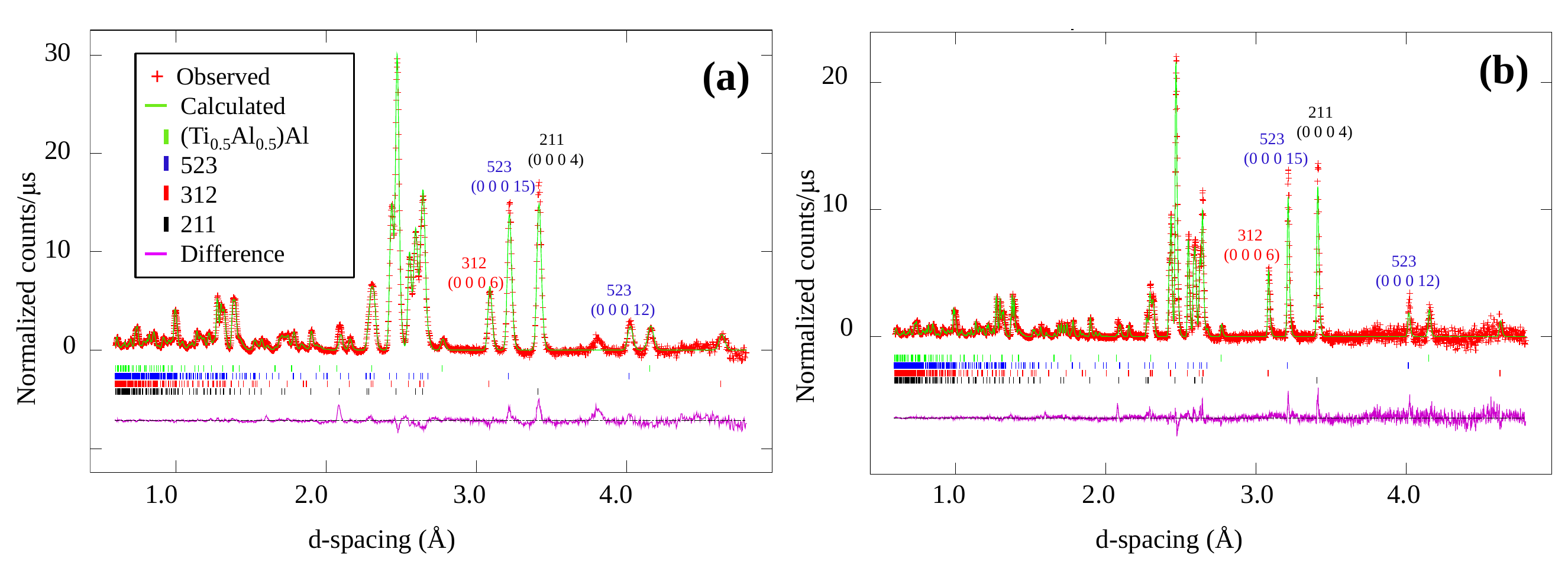}
\caption{
Rietveld analysis of neutron diffraction data measured on HIPPO at 100$\deg$C for muliphase sample from (a) 90$\deg$ 
detector bank and (b) 144$\deg$ detector bank. Raw data points are shown as red + symbols and 
the calculated profile is shown as a solid green line. Underneath, markers show calculated peak 
positions of each phase. From top to bottom: (Ti$_{0.5}$Al$_{0.5}$)Al (green), 
523 (blue), 312
(red), and 211 (black). Difference curve 
($Y_\mathrm{obs}-Y_\mathrm{calc}$) is shown in bottom of each panel as a solid purple line.
}
\label{fig:523rietveld}
\end{figure*}

\subsection{Structure refinement}

A texture analysis using the Material Analysis Using Diffraction, MAUD, code 
\cite{MAUD2010,HIPPO2} showed a mild 
fiber texture for all temperature points. As the texture was weak, the data from 
the three rotations and the detector rings were integrated, resulting in one 
histogram per detector bank. Due to the large detector coverage of HIPPO, this 
procedure is similar to spinning the sample to randomize the preferred 
orientation. The neutron time-of-flight data were analyzed as texture-free powders 
with the Rietveld method using the General Structure Analysis System, GSAS \cite{GSAS}.
Only the higher resolution data from the 90$\deg$, 120$\deg$, and 144$\deg$ 
banks were used in the analysis. 
The MAUD refinements, which incorporated preferred orientation, 
gave parameters that were within 
error bars of those determined without including texture by Rietveld refinement with GSAS. 
Therefore, all results reported herein are from the GSAS refinements assuming 
random texture.

The \textit{gsaslanguage} refinement script 
 \cite{gsaslang} was used to ensure that identical refinement strategies 
were employed for all temperatures. The instrument alignment (DIFC parameter 
in GSAS) was fixed for the backscattered (144$\deg$) detector bank, which has the 
highest resolution, and refined for the 90$\deg$ and 120$\deg$ bank for the 
lowest temperature run  at 100$\deg$C. For subsequent runs, DIFC was fixed for 
all three banks. 
Refined parameters were 16 background parameters of GSAS background function 
\#1, lattice parameters of all phases, phase fractions, $\sigma_1$ profile parameter for peak 
width, atomic positions, and isotropic thermal motion parameters.
The atomic positions in Ti$_2$AlC [$z$ coordinate of Ti;
 see Fig. \ref{fig:523struct}(a)] and in 
Ti$_3$AlC$_2$ [$z$ coordinates of Ti$_\mathrm{I}$ and C; see Fig. \ref{fig:523struct}(b)] were 
both refined in the $P6_3/mmc$ space group. 

For Ti$_5$Al$_2$C$_3$, there are 
six unique atomic sites: Ti$_\mathrm{I}$, Ti$_\mathrm{II}$, and 
C$_\mathrm{I}$ with 312-like stacking, Ti$_\mathrm{III}$ and 
C$_\mathrm{II}$ with 211-like stacking, and one Al site between 211- and 
312-stacked octahedra throughout the cell [see Fig. \ref{fig:523struct}(c)]. 
The best space group to 
represent these positions was found to be $R \bar{3} m$; therefore, the $z$ coordinates of 
Ti$_\mathrm{I}$, Ti$_\mathrm{III}$, Al, and C$_\mathrm{I}$ were refined 
according to the constraints induced by the trigonal $R \bar{3} m$ space group. 

In addition to the three MAX phase carbides, peaks 
corresponding to an intermetallic that crystallizes like $\gamma$-TiAl, with 
tetragonal space group $P_4/mmm$ \cite{TiAl}, were found. Refinement of the site occupancy 
factor on the Ti site led to 50\% Ti and 50\% Al antisite defects, yielding a 
stoichiometry of (Ti$_{0.5}$Al$_{0.5}$)Al. Given the nearly null scattering
intensity of the mixed site, and the low phase fraction of this phase, the isotropic thermal 
motion parameters were constrained together to reduce the number of variables.

\subsection{First-principles calculations}

First-principles phonon calculations were used to calculate the anisotropic mean-squared atomic displacements. The DFT calculations were performed using the projector-augmented wave (PAW) method \cite{PAW}, as implemented in the Vienna Ab initio Simulation Package (VASP) code \cite{vasp1,vasp2,vasp3}. The generalized gradient approximation (GGA) exchange-correlation functional of Perdew-Burke-Ernzerhof (PBE) was used \cite{PBE} with a cutoff of 500 eV. The total energy was converged to 10$^{-8}$ eV, with a $k$-point grid of $12 \times 12 \times 4$ for Ti$_2$AlC and Ti$_3$AlC$_2$ and $12\times 12 \times 2$ for T$_5$Al$_2$C$_3$.

The real-space force constants were calculated using density functional perturbation theory (DFPT) \cite{DFPT} implemented in the VASP code. The phonon frequencies were calculated from the force constants with Phonopy \cite{phonopy}, and the ADPs were calculated from
the frequencies and eigenvectors of the force constant matrix. Further details on the ADP calculations can be found in Ref. \cite{adpmax}.

\section{Results}

\begin{figure}
\centering
		\includegraphics[width=0.75\columnwidth]{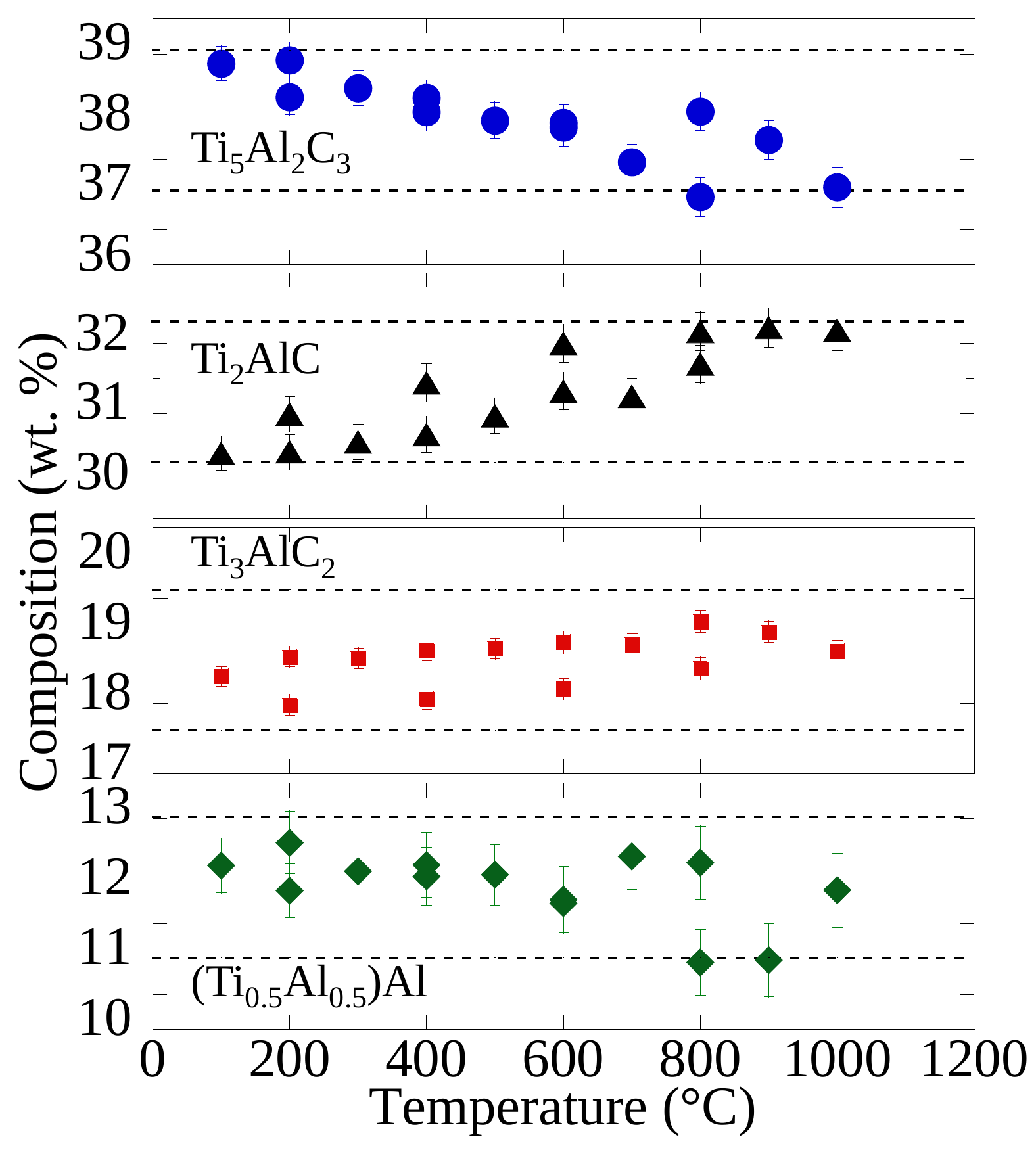}
\caption{
Compositions (wt.\%) of each phase in the Ti--Al--C sample as a function of temperature 
upon heating and cooling. Dashed lines indicate limits for $\pm$1\% range.
}
\label{fig:523wt}
\end{figure}

\begin{table}
\caption
{Profile agreement factors {rfactors} for Rietveld refinements of neutron diffraction data collected during heating and cooling for the multiphase Ti--Al--C sample.}
	\label{table:523paf}
\centering
		\begin{tabular}{l ccc}
\hline \hline
T ($\deg$C)	& $wR_p$ (\%) & $\chi^2$ & $R_\mathrm{exp}$ \\ 
\hline
100 &	1.34	& 3.959 &	0.67 \\
200 &	1.31	& 4.042 &	0.65 \\ 
600 &	1.23	& 3.599 &	0.65 \\
1000 &	1.17 &	3.331 &	0.64 \\
600$^\dagger$ &	1.29 &	3.998 &	0.65 \\
200$^\dagger$  &	1.4	& 4.697 &	0.65 \\
\hline \hline
\end{tabular}
\flushleft $^\dagger$ Data collected during cooling.
\end{table}

As noted above, texture analysis (Fig. \ref{fig:523texture}) 
showed a mild (000$l$) fiber texture for all 
three MAX phases. Not surprisingly, the texture did not change during heating 
or cooling [compare Figs. \ref{fig:523texture}(a), (b), (c) to Figs. \ref{fig:523texture}(d), (e), (f), respectively]. 
The Rietveld fits for the neutron time-of-flight data -- integrated for the full detector rings and the 
three measured orientations -- are shown for the lowest temperature run at 100$\deg$C for the 
90$\deg$ [Fig. \ref{fig:523rietveld}(a)] and the 144$\deg$ detector banks [Fig. 
\ref{fig:523rietveld}(b)]. The calculated fit (solid green lines) and measured data (red plus signs) 
are compared, with the difference curve plotted at the bottom (solid purple line). The markers 
above the difference curve show the peak positions for the phases: from top to bottom, 
(Ti$_{0.5}$Al$_{0.5}$)Al (green), 523 (blue), 312 (red), and 
211 (black). The higher-d-spacing peaks resulting from diffraction by the basal planes are 
labeled for the (0 0 0 6) peak of Ti$_3$AlC$_2$, the (0 0 0 4) peak of Ti$_2$AlC, and 
the (0 0 0 15) and (0 0 0 12) peaks of Ti$_5$Al$_2$C$_3$. The latter two 
peaks unambiguously identify Ti$_5$Al$_2$C$_3$ as the dominant phase, 
as they cannot be accounted for by 
any other known phase in the Ti--Al--C system.  

The profile agreement factors for the Rietveld fits 
are listed in Table \ref{table:523paf} at select temperatures, 
giving the weighted profile $R$ index, $wR_p$, 
the goodness of fit, $\chi^2$, and the expected $R$ factor, $R_\mathrm{exp}$ \cite{rfactors}. 
Good agreement is observed between the calculated and observed profiles, with one unidentified 
peak at 2.09 \AA~ whose origin remains unclear. A broad peak at 3.80 \AA~ is present only in the 90$\deg$ bank [Fig. \ref{fig:523rietveld}(a)], which is likely due to background interference.

The composition determined from Rietveld analysis is 
 $38(\pm 1)$ wt. \% Ti$_5$Al$_2$C$_3$, $32(\pm 1)$ wt.\% Ti$_2$AlC, $18(\pm 
1)$ wt. \% Ti$_3$AlC$_2$, and $12(\pm 1)$ wt.\% (Ti$_{0.5}$Al$_{0.5}$)Al. 
The temperature dependencies of the fractions of each phase are plotted in Fig. 4, where the 
dashed lines indicate the $\pm$1 wt.\% limits. The compositions generally stay within 1 wt.\% of 
the average value during heating and cooling, lending credibility to our data analysis and the 
resulting uncertainties for the compositions. 

\begin{figure*}
\centering
		\includegraphics[width=0.90 \textwidth]{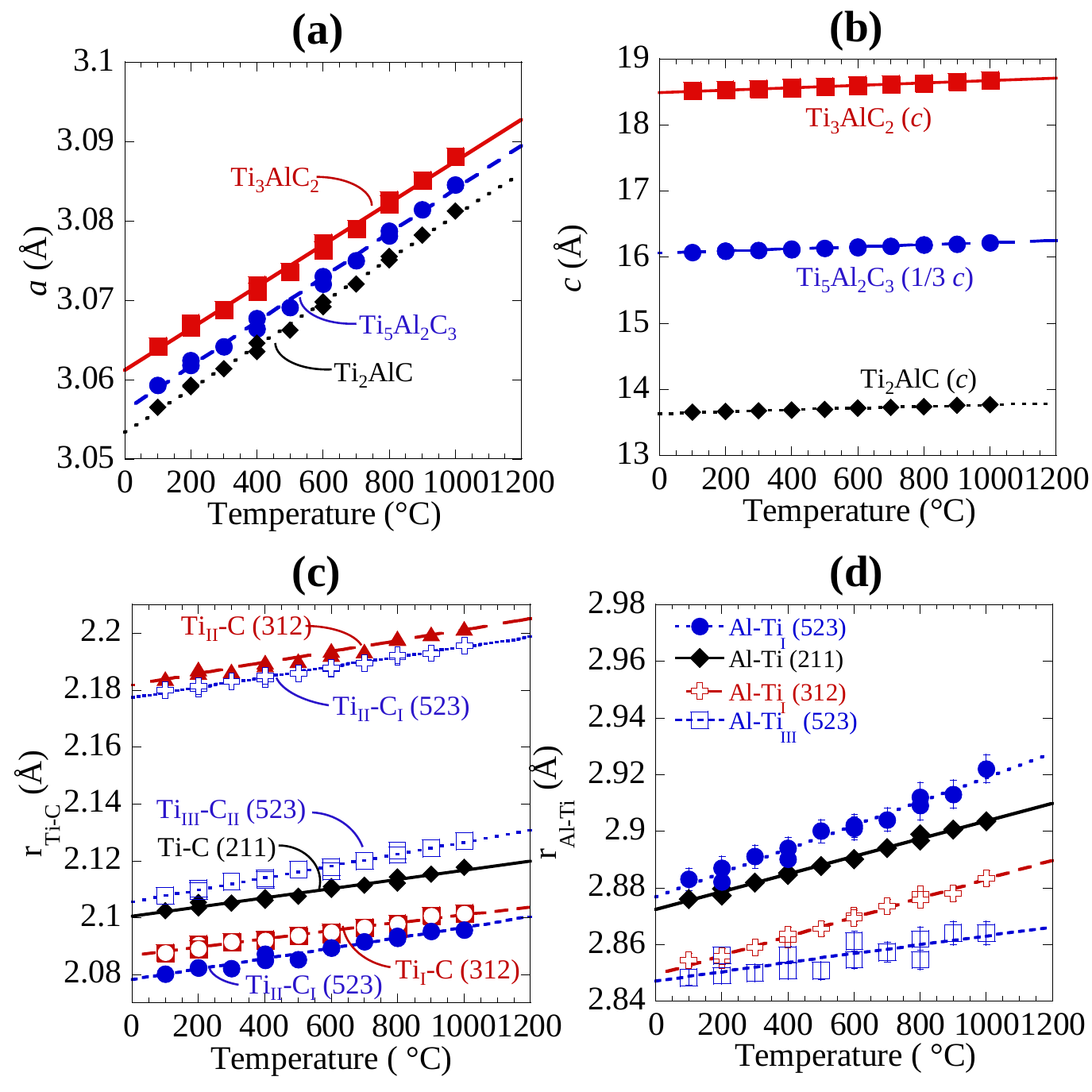}
\caption{
Temperature dependence of (a) $a$ and (b) $c$ lattice parameters, and (c) Ti--C and (d) Ti--Al interatomic distances in Ti$_5$Al$_2$C$_3$ (blue), Ti$_3$AlC$_2$ (red) and Ti$_2$AlC (black). Refer to Fig. \ref{fig:523struct} for notations of Ti and C atoms. Errors are typically smaller than symbol size.}
\label{fig:523bonds}
\end{figure*}

\begin{figure*}
\centering
		\includegraphics[width= 0.85 \textwidth]{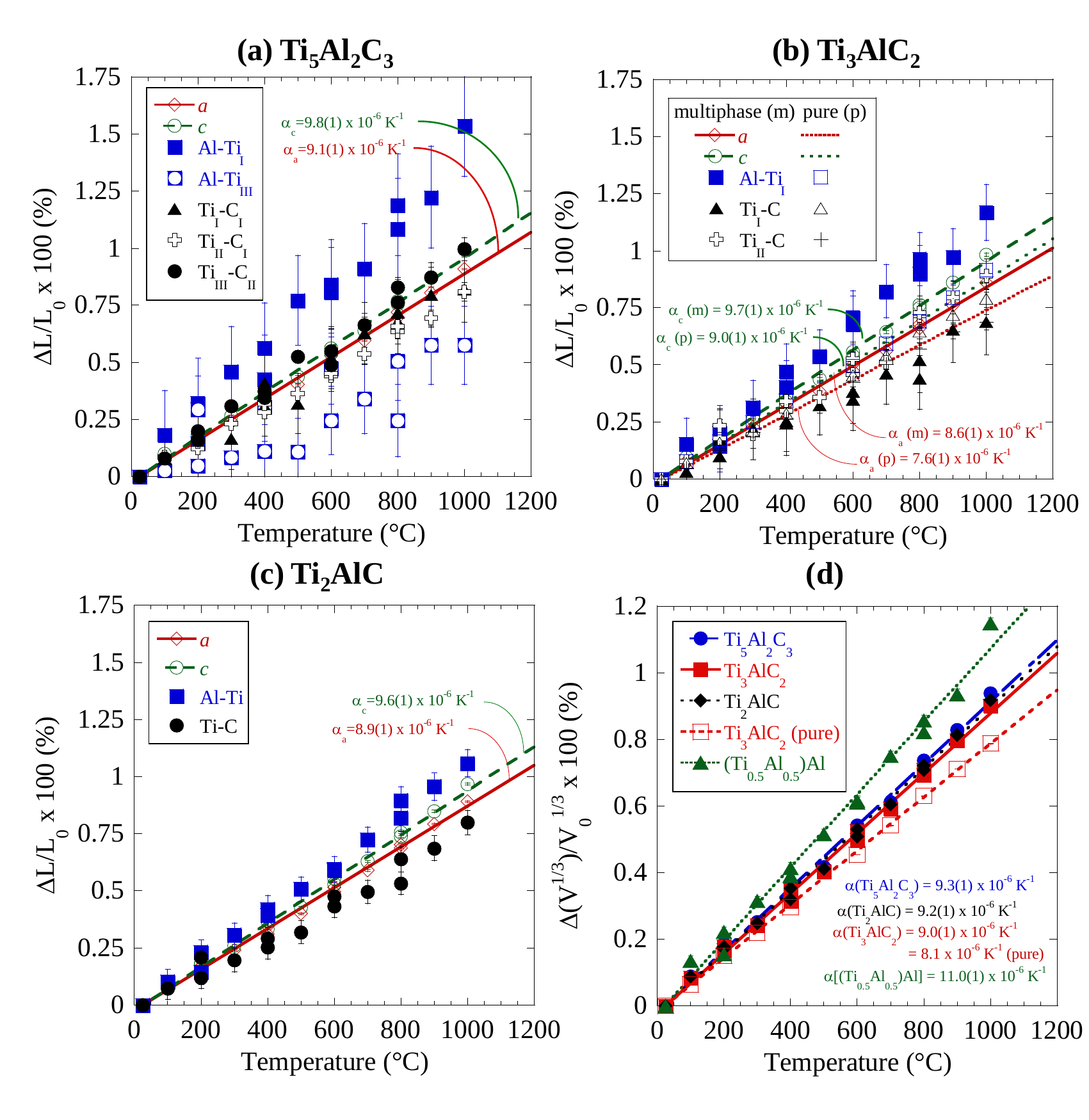}
\caption{
Temperature dependences of the thermal strains of the lattice parameters and interatomic distances in (a) Ti$_5$Al$_2$C$_3$, (b) Ti$_3$AlC$_2$, and (c) Ti$_2$AlC for the multiphase sample. In (b), the results for predominantly single-phase Ti$_3$AlC$_2$ are also shown. (d) Volume expansion of Ti$_5$Al$_2$C$_3$ (blue circles), Ti$_3$AlC$_2$ (red solid squares), Ti$_2$AlC (black diamonds), and (Ti$_{0.5}$Al$_{0.5}$)Al (green triangles) in the multiphase sample, and pure Ti$_3$AlC$_2$ (open red squares). Room temperature (25$\deg$C) values were extrapolated to use for $L_0$ and $V_0$. Errors for volume expansions in (d) are typically smaller than symbol size. \vspace{-1em}}
\label{fig:523exp}
\end{figure*}

The temperature-dependent expansions of the lattice parameters and interatomic distances in 
Ti$_5$Al$_2$C$_3$, Ti$_3$AlC$_2$, and Ti$_2$AlC are shown in Figs. \ref{fig:523exp}(a), 
(b), and (c), respectively.   
In Fig. \ref{fig:523exp}(b), the parameters for the predominantly 
single-phase Ti$_3$AlC$_2$ sample are also plotted as open symbols.
The anisotropic TEC values are listed in Table \ref{table:523exp}, along with those from previous studies 
on Ti$_2$AlC \cite{2000Ti2AlN-C} and Ti$_3$AlC$_2$ \cite{Pang312,ScabaroziCTE,Tzenov}.

To further compare the thermal expansions of the three phases, 
the temperature dependencies of $(\Delta V/V_o)^{1/3}$ -- where $\Delta V$ is the
change in unit cell at temperature $T$ as compared to that at the 
reference temperature, 25$\deg$C (extrapolated), $V_0$ -- are plotted in Fig \ref{fig:523exp}(d).  The results for (Ti$_{0.5}$Al$_{0.5}$)Al are also shown.  The slope of these lines yields the average thermal expansion $\alpha_\mathrm{av}$. 
From these results we find that for Ti$_2$AlC, 
$\alpha_\mathrm{av} = 9.2(\pm 0.1)$ \ex; for Ti$_3$AlC$_2$, 
$\alpha_\mathrm{av} = 9.0 (\pm 0.1)$ \ex; for Ti$_5$Al$_2$C$_3$
 $\alpha_\mathrm{av} = 9.3 (\pm 0.1)$ \ex. It is thus clear from Fig. \ref{fig:523exp}(d)
that the TECs of the three MAX phases are almost identical within 
the error bars.

\begin{table*}
\caption
{Thermal expansions from HTND for Ti--Al--C phases in the sample studied in this work, along with those from other studies 
\cite{Pang312,2000Ti2AlN-C,ScabaroziCTE,Tzenov,HahnTiAl,HeTiAl} determined through HTXRD and dilatometry. Numbers in parentheses are estimated standard deviations in the last significant digit of the refined parameters.}
	\label{table:523exp}
\centering
		\begin{tabular}{p{8 em}lllll}
\hline \hline
Phase &	$\alpha_a$ 	& $\alpha_c$ & $\alpha_\mathrm{av}$ & Anisotropy & Ref.\\ 
& ($10^{-6}$ K$^{-1}$) & ($10^{-6}$ K$^{-1}$)  & ($10^{-6}$ K$^{-1}$) & ($\alpha_c/\alpha_a$) & \\
\hline
Ti$_5$Al$_2$C$_3$ & 
9.1(1)$^\dagger$ & 9.8(1)$^\dagger$ & 9.3(1)$^\dagger$ &	1.08(3)$^\dagger$ & 	
This work \\
\hline Ti$_2$AlC &	
9.0(1)$^\dagger$ &	9.6(1)$^\dagger$ & 9.2(1)$^\dagger$ & 1.07(3)$^\dagger$ &
This work \\
&
7.1(3)$^a$ & 10.0(5)$^a$  & 8.1(5)$^{a,b}$; 8.2(2)$^c$ & 1.41(4) &
Ref. \cite{2000Ti2AlN-C} \\
\hline Ti$_3$AlC$_2$ &
8.6(1)$^\dagger$ & 9.7(1)$^\dagger$ & 9.0(1)$^\dagger$ & 1.13(3)$^\dagger$ & 
This work \\
&
7.6(1) & 9.0(1) & 8.1(1) & 1.18(3) & 
This work \\
&
8.3(1)$^a$ & 11.1(1)$^a$ & 9.2(1)$^{a,b}$; 7.9(5)$^c$ & 1.33(1)$^a$ &
Ref. \cite{ScabaroziCTE} \\
&
- &	- & 9.0(2)$^c$ &	- &
Ref. \cite{Tzenov} \\
&
8.5	& 10.2 & 9.2$^b$ & 1.2 & 
Ref. \cite{Pang312} \\
\hline (Ti$_{0.5}$Al$_{0.5}$)Al &	 
10.7(1)$^\dagger$ & 11.5(1)$^\dagger$ &	11.0(1) $^\dagger$ &	 1.074(3)$^\dagger$ &
This work \\
$\gamma$-TiAl &
- & - & 10.0 & - &
Ref. \cite{HahnTiAl} \\
&
9.77 &	9.26	& - & - &
Ref. \cite{HeTiAl} \\
\hline \hline
\end{tabular}
\flushleft 
$^\dagger$ Multiphase sample. \\
$^a$ High-temperature XRD. \\
$^b$ Assuming $\alpha_\mathrm{av} = (2/3 \alpha_a+1/3 \alpha_c) = V_0^{-1/3}dV^{1/3}/dT$. \\
$^c$ Dilatometry.
\end{table*}

The absolute values of the $c$ and $a$ lattice parameters 
[Figs. \ref{fig:523bonds}(a) and (b), Table \ref{table:523lp}] 
are also comparable, but it is apparent that Ti$_3$AlC$_2$ has the highest $a$ lattice 
parameter and Ti$_2$AlC has the lowest, with that of Ti$_5$Al$_2$C$_3$ falling in between. The same is true of the $c$ lattice parameters, after normalizing it by 
three to account for the three formula units in Ti$_5$Al$_2$C$_3$ [Fig. \ref{fig:523bonds}(b)].  

The temperature dependences of the absolute values of the Ti--C and Al--Ti bonds are shown in Figs. \ref{fig:523bonds}(c) and (d), respectively. The bonds in 211, 312, and 523 are shown in black, red, and blue, respectively. Note that the absolute range for the scale is the same for the graphs shown in Figs. \ref{fig:523bonds}(c) and \ref{fig:523bonds}(d). The extrapolated room temperature values for all bonds are also listed in Table \ref{table:523bonds}, along with their expansion rates.  

Due to the overlap of peaks in our data set we were unable to determine the anisotropic displacements as was done in our previous HTND studies \cite{adpmax,312paper,211paper,Claudia1999,ThermalTi4AlN3}.  Instead, we examine the isotropic ADPs, $U_\mathrm{iso}$, which represent the mean-squared displacements of the atoms from their equilibrium positions. Figures \ref{fig:523adp}(a), (b), and (c) show the temperature dependences of $U_\mathrm{iso}$ for the unique Ti (red), Al (blue), and C (black) atoms in Ti$_5$Al$_2$C$_3$, Ti$_3$AlC$_2$, and Ti$_2$AlC, respectively. In Fig. \ref{fig:523adp}(b) the values for predominantly single-phase Ti$_3$AlC$_2$ are shown for comparison's sake.

The values calculated with first-principles phonon calculations are shown as lines. In both experimental and calculated results, the Al atom shows the highest amplitude. Figure \ref{fig:U-Al-all} compares the experimental and calculated values of $U_\mathrm{iso}$ for the Al atoms in all three phases Ti$_5$Al$_2$C$_3$, Ti$_3$AlC$_2$ and Ti$_2$AlC in the multiphase sample, along with $U_\mathrm{iso}$ for pure Ti$_3$AlC$_2$. Also shown are the experimental $U_\mathrm{eq}$ values for the Al-containing phases Ti$_2$AlN and Ti$_4$AlN$_3$, determined from in previous HTND studies \cite{211paper,ThermalTi4AlN3}, along with calculated values for those phases from Ref. \cite{adpmax}. The $U_\mathrm{iso}$ values for Al in the three Ti--Al--C phases are similar to each other, and slightly higher than $U_\mathrm{eq}$ of for Al in the Ti--Al--N phases, which is consistent with the values determined from first-principles calculations.  Note that the calculated $U_\mathrm{iso}$ curves for Al in Ti$_5$Al$_2$C$_3$, Ti$_2$AlC, and Ti$_3$AlC$_2$ in Fig. \ref{fig:523adp} lie essentially on top of one another, and those for Al in Ti$_2$AlN and Ti$_4$AlN$_3$ are similar as well, but smaller than those of the carbide phases.

\section{Discussion}

\subsection{Lattice parameters, expansions, and anisotropies}

%
Not surprising, the measured $a$- and scaled $c$-parameters of the 523 phase are in between those of the 211 and 312 phases [Fig. \ref{fig:523bonds}(b)].  This is also consistent with the values from our first-principles calculations (Table \ref{table:523lp}). Our lattice parameters are in good agreement with literature values for Ti$_2$AlC \cite{Ti2AlN-CXray,2000Ti2AlN-C,523paper}, Ti$_5$Al$_2$C$_3$ \cite{523paper}, and Ti$_3$AlC$_2$ \cite{PietzkaSchuster,523paper}.
Their order of increase is likely due to intricacies in charge transfer involved in bonding.  The fact that the $a$ lattice parameter scales with the number of Ti--C bonds is consistent with first principles calculations \cite{523paper}.

The overall expansions of the three MAX phases in the multiphase sample are, within their error bars, nearly equivalent (Table \ref{table:523exp}). The $a$ and $c$ lattice expansions are qualitatively comparable, with the $a$ lattice parameters and their thermal expansions all within 1\% of each other [Fig. \ref{fig:523bonds}(a)].  Consistent with previous studies, the expansion in the $c$ direction is greater than along the $a$.  However, for reasons discussed below, in the present study the degree of anisotropy is lower than in Refs. \cite{2000Ti2AlN-C,ScabaroziCTE,Tzenov} (Table \ref{table:523exp}). We now consider each of the phases separately. \\

\paragraph*{Ti$_2$AlC:}

 The TECs along the $a$- and $c$-direction for the Ti$_2$AlC sample measured herein --
9.6($\pm 0.1$) \ex and 8.9($\pm 0.1$) \ex respectively [Fig. \ref{fig:523exp}(c)] -- fall in between those reported previously for Ti$_2$AlC (Ref. \cite{2000Ti2AlN-C}). The reason(s) for the discrepancy is unknown at this time but could very well reflect differences in chemistry. Recent work in the literature suggests that Ti$_2$AlC exists over a range of stoichiometries. For example, Bai \textit{et al.} recently reported the existence of a Ti$_2$AlC$_x$ phase where $x$ was as low as 0.69 \cite{bai2009rapid,bai2012microstructures}. Herein, it is more likely than not that the Ti$_2$AlC is Al-deficient since it is believed that the loss of Al is what triggers the transformation to the 523 and possibly the 312 phase. 

\paragraph*{Ti$_3$AlC$_2$:} 

The TEC values measured herein for the 312 phase depended on sample.  The predominantly single-phase Ti$_3$AlC$_2$ sample has a lower expansion in both directions, resulting in a statistically significant lower $\alpha_\mathrm{av}$ of $8.1(\pm 0.1)$ \ex (Fig. \ref{fig:523exp}). 
At 9.0($\pm 0.1$) \ex,
$\alpha_\mathrm{av}$ for the 312 phase in the multiphase sample is about 10\% higher than in the single phase one. 

\paragraph*{Ti$_5$Al$_2$C$_3$:}

Since this is the first report on the effect of temperature on the lattice parameters of the 523 phase, there are no previous results to compare them with. However, the fact that $\alpha_\mathrm{av}$ of this phase is very comparable to the 211 and 312 phases is not surprising and is consistent with the fact that the former is comprised of the same building blocks as the latter. \\

\begin{figure*}
\centering
	\includegraphics[width= \textwidth]{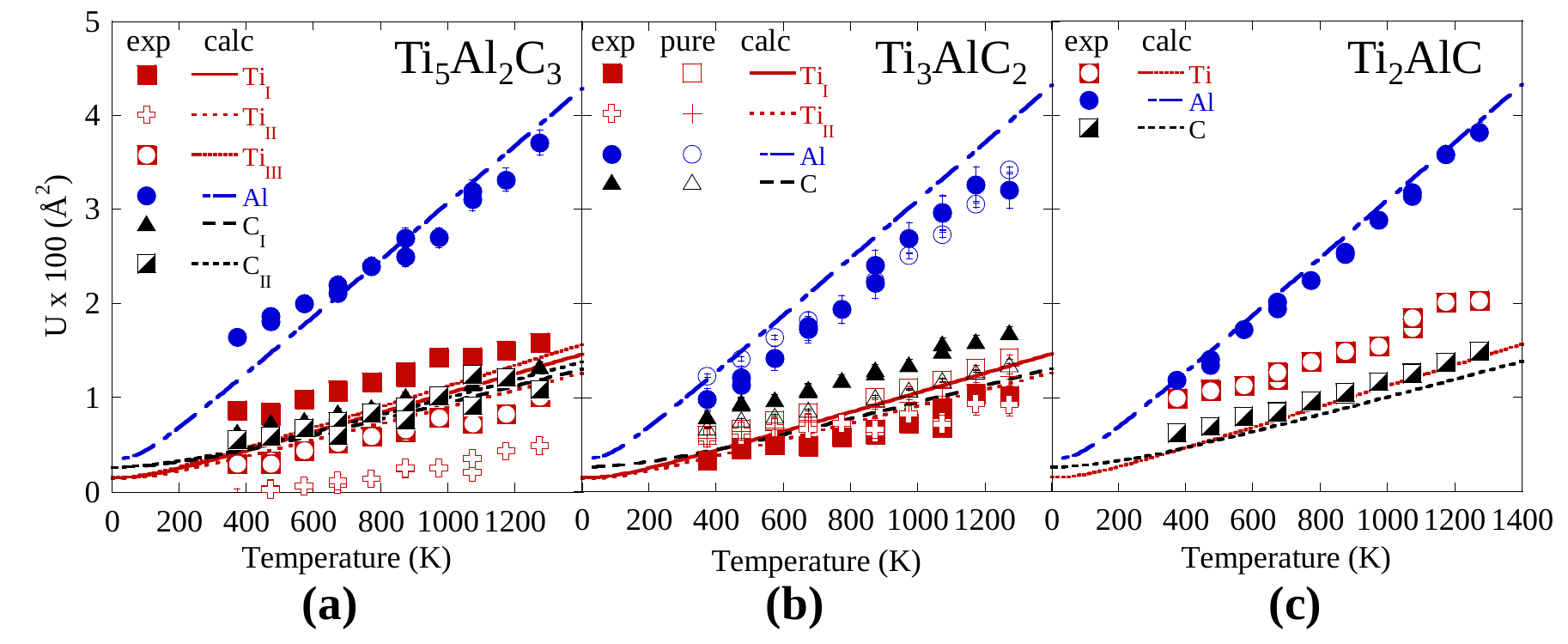}
\caption{
Temperature dependence of isotropic thermal displacement parameter $U_\mathrm{iso}$ during heating and cooling for atoms in (a) Ti$_5$Al$_2$C$_3$, (b) Ti$_3$AlC$_2$, and (c) Ti$_2$AlC.  Errors are typically smaller than symbol size.}
\label{fig:523adp}
\end{figure*}

Lastly, a few remarks on the expansions. The ability to measure phase sensitive TECs is an advantage of HTND, as compared to other methods such as dilatometry, that require pure phases to measure their volume TECs. However, it is important to appreciate that the TEC values measured herein per force are less anisotropic than those one would measure in loose powders. In the latter case, the solid is free to expand, whereas when the measurement is made on bulk solids, residual stresses can accrue and reduce the values of the thermal expansions in various directions. The effect is best appreciated when the TECs in the $a$ and $c$ directions are compared with those measured on powder Ti$_2$AlC samples \cite{ScabaroziCTE}.

\begin{table*}
\caption
{Temperature-dependent $a$ and $c$ lattice parameters from Rietveld refinement of neutron diffraction data collected during heating and cooling.  Numbers in parentheses are estimated standard deviations in the last significant figure of the refined parameters.  Room temperature values are extrapolated to 25$\deg$C from linear interpolation.  
}
	\label{table:523lp}
\centering
		\begin{tabular}{p{8em}l p{6em} lp{6em} ll}
\hline \hline
 & \multicolumn{2}{c}{Ti$_5$Al$_2$C$_3$} &
\multicolumn{2}{c}{Ti$_2$AlC} & \multicolumn{2}{c}{Ti$_3$AlC$_2$} \\
Temp. ($\deg$C) & $a$ (\AA) & $c$ (\AA) & $a$ (\AA) & $c$ (\AA)	& $a$ (\AA)	& $c$ (\AA) \\
\hline
RT (Ref.\cite{PietzkaSchuster}) &	-&	-&	-&	-&	3.0753	&18.578 \\
RT (Ref. \cite{Ti2AlN-CXray}) &-&	-&	3.065(4)	& 13.71(3)&	-&	- \\
RT (Ref. \cite{2000Ti2AlN-C})&-&	-&	3.051&	13.637&	-&	- \\
RT (Ref. \cite{523paper}) &	3.064(2) & 48.23(2) & 3.063	&13.645	&3.060	&18.66 \\
RT (Ref. \cite{523paper})$^a$  & 3.068 & 48.186 & 3.067 & 13.75 & 3.083 & 18.661 \\
RT$^b$ & 3.05678	& 48.189	& 3.05405	& 13.6422	& 3.06186	& 18.4994 \\
100 & 	3.05926(6)	& 48.237(1)	& 3.05656(7)	& 13.6551(5)	& 3.06424(8)	& 18.5172(8)\\
200& 	3.06184(6)& 	48.280(1)& 	3.05912(7)	& 13.6669(4)	& 3.06667(8)	& 18.5331(8)\\
300	&  3.06419(6)	& 48.317(1)& 	3.06141(7)& 	13.6777(5)& 	3.06887(8) & 18.5483(8)\\
400& 	3.06642(6) & 48.354(1) & 3.06353(7) & 13.6879(4) & 3.07108(8) &  18.5617(8)\\
500 & 3.06911(6)& 	48.400(1)& 	3.06627(7) & 13.7006(4)	& 3.07359(8)	& 18.5803(8)\\
600	& 3.07208(6)	& 48.449(1)	& 3.06920(7)	& 13.7145(4)	& 3.07635(8)	& 18.5991(8)\\
700	& 3.07501(6)	& 48.497(1)	& 3.07206(7)	& 13.7280(4)	& 3.07908(8)	& 18.6184(8)\\
800	& 3.07814(6)	& 48.554(1)	& 3.07510(7)	& 13.7431(5)	& 3.08213(8)	& 18.6380(8)\\
900	& 3.08141(6)	& 48.608(2)	& 3.07825(7)	& 13.7581(5)	& 3.08519(8)	& 18.6587(9)\\
1000	& 3.08460(6)	& 48.667(2)	& 3.08126(7)	& 13.7743(5)	& 3.08818(8)	& 18.6811(9)\\
800$^c$	& 3.07876(6)	& 48.559(2)	& 3.07552(7)	& 13.7453(5)	& 3.08264(9)	& 18.6404(9)\\
600$^c$	& 3.07301(6)	& 48.459(2)	& 3.06978(7)	& 13.7180(5)	& 3.07730(9)	& 18.6025(9)\\
400$^c$	& 3.06770(6)	& 48.368(2)	& 3.06460(7)	& 13.6922(5)	& 3.07200(9)	& 18.5684(9)\\
200$^c$	& 3.06239(6)	& 48.279(2)	& 3.05928(7)	& 13.6672(5)	& 3.06712(9)	& 18.5337(9)\\
\hline \hline
\end{tabular}
\flushleft 
$^a$ DFT calculations.\\
$^b$ Extrapolated value. \\
$^c$ Data collected during cooling. \\
\end{table*}

\begin{table}[t]
\caption
{Interatomic distances in Ti$_5$AlC$_2$, Ti$_2$AlC, and Ti$_3$AlC$_2$ in 
the multiphase sample from Rietveld refinement of neutron diffraction 
data collected during heating and cooling,
along with their expansions.  All values are extrapolated to 25$\deg$C from linear interpolation.  Numbers in parentheses are estimated standard deviations in the last significant digit of the refined parameters.
}
	\label{table:523bonds}
\centering
		\begin{tabular}{llll}
\hline \hline
Phase & Bond & Bond length & Bond expansion$^\dagger$  \\
           &              & (\AA)          & ($10^{-6}$ K$^{-1}$) \\
\hline
Ti$_5$Al$_2$C$_3$ & Al--Ti$_\mathrm{I}$ & 2.878  & 14.6 \\
& Al--Ti$_\mathrm{III}$ & 2.848 & 5.5 \\
& Ti$_\mathrm{I}$--C$_\mathrm{I}$ & 2.079 & 8.8 \\
& Ti$_\mathrm{II}$--C$_\mathrm{I}$ & 2.180 & 8.1 \\
& Ti$_\mathrm{III}$--C$_\mathrm{III}$ & 2.106 & 10.0 \\
\hline
Ti$_2$AlC & Al--Ti & 2.873 & 10.8 \\
& Ti--C & 2.101 & 7.7 \\
\hline
Ti$_3$AlC$_2$ & Al--Ti$_\mathrm{I}$ & 2.854 & 11.8 \\
& Ti$_\mathrm{I}$--C & 2.087  & 6.8 \\
& Ti$_\mathrm{II}$--C & 2.182 & 8.8 \\
\hline \hline
\end{tabular}
\flushleft 
$^\dagger$ Bond expansion: $L^{-1} \cdot dL/dT$ from least-squares fit of $\Delta L/L_0$ vs. $T$.
\end{table}

\subsection{Bond lengths}

While the overall expansions and anisotropies in the three MAX-like phases are comparable, the most interesting aspect of this work is the relationship between bond length evolution and the stacking of the octahedra among the three phases.  In the literature, it is fairly well established, both experimentally and theoretically, that the $M$--C bonds adjacent to the $A$ layers (\textit{i.e.} Ti$_\mathrm{I}$-C in Fig. \ref{fig:523struct}) are shorter than those in the stoichiometric binary $MX$, while the ones that are not, (\textit{viz.} Ti$_\mathrm{II}$--C, in Fig. \ref{fig:523struct}) are longer. Figure \ref{fig:523bonds}(c) and the Ti--C lengths in Table \ref{table:523bonds} are fully consistent with this general conclusion. Not surprisingly, the Ti--C bond lengths in the 211 slab in the 523 phase are almost identical to those of the 211 phase [Fig. \ref{fig:523bonds}(c)]. Similarly, the Ti--C bond lengths in the 312 phase are very similar to those of the 312 slabs in the 523 phase [Fig. \ref{fig:523bonds}(c)]. This applies not only to the absolute Ti--C bond lengths values but also to their thermal expansions, which are quite comparable as well [Fig. \ref{fig:523struct}(c)].   It should be noted that the longest bonds in the 312-stacked octahedra in Ti$_5$Al$_2$C$_3$ are slightly shorter than $r_\mathrm{Ti_{II}-C}$ in Ti$_3$AlC$_2$, while the Ti--C bonds in the 211-stacked octahedra in Ti$_5$Al$_2$C$_3$ are slightly longer than those in Ti$_2$AlC [Fig. \ref{fig:523bonds}(c)].  This suggests that the structure is slightly more uniform than the individual 211- and 312-stacked phases due to the interleaved nature of the stacking sequences.  These comments notwithstanding, it is clear from Fig. \ref{fig:523bonds}(c) that the same structural units behave similarly.  These results are gratifying because they indirectly validate our Rietveld analysis.

The situation for the Ti-Al bonds is not as clear.  
Since the difference between the 211 and 312 phases is the number of Ti--C octahedra between Al layers, it is expected that only the Ti--C bonds would be significantly affected by the change in Ti--C stacking while the Al--Ti bonds should be similar among the three phases.  However, we find that the Al--Ti bonds are clearly affected by stoichiometry [Fig. \ref{fig:523bonds}(d)]: $r_\mathrm{Ti-Al}$ is significantly longer in the 211 phase [black diamonds in Fig. \ref{fig:523bonds}(d)] than in the 312 phase [red crosses in Fig. \ref{fig:523bonds}(d)], while the opposite is true of those bonds in the 523 phase; \textit{i.e.}, $r_\mathrm{Al-Ti_I}> r_\mathrm{Al-Ti_{III}}$ in Ti$_5$Al$_2$C$_3$ [compare blue circles and blue squares in \ref{fig:523bonds}(d)]. 
The reasons for this state of affairs are not fully understood, but are likely related to the following observations:

\begin{enumerate}[(i)]
\item The Ti--C bonds are relatively stiff building blocks of the individual 312 and 211 units, as evidenced by the fact that they stay relatively the same size as in the original Ti$_3$AlC$_2$ and Ti$_2$AlC phases when the stacking sequences are interleaved.  The dimension within the crystal that thus has the most flexibility is the Al--Ti bond.  Therefore, it is likely that the  Al--Ti bond plays a role as an effective ``compensating spring'' in the structure to minimize the crystal energy.  Furthermore, this role would be different -- and most probably more dominant -- in the more complex  Ti$_5$Al$_2$C$_3$ higher-order phase.
\item Among the possible factors that could be compensated for in the flexible Al--Ti bond discussed above are those related to constraints on the lattice parameters -- especially on the $c$-lattice parameter, which essentially determines the Al--Ti bond length, given that the Ti--C octahedra are rigid blocks.  In a sample with multiple competing phases, it is likely that these effects are prominent and manifest themselves in the Al--Ti bond.
\item The Al atoms in Ti$_2$AlC and Ti$_3$AlC$_2$ lie in a mirror plane within the structures, while Al is not constrained to mirror symmetry between the Ti--C atoms in the Ti$_5$Al$_2$C$_3$ phase (see Fig. \ref{fig:523struct}).  Therefore, the changes in the Al--Ti distances in the 211-stacked and 312-stacked structures that occur when they are interleaved to form 523 
may be a consequence of the symmetry break.
\end{enumerate}

 These comments notwithstanding, it is important to note that the average Ti--Al bond length in 523 (2.863 \AA) is equal the average of the Al--Ti bond lengths in 312 and 211 (also 2.863 \AA). The average expansions of those bonds are also similar (see below).  While more work is needed to fully understand the Al--Ti bond length behavior, 
it can be reasonably concluded that the dimensions of the Ti--C units are consistent for a given stacking,
regardless of whether they are interleaved in a higher-order phase or in a conventional
MAX phase.  Based on this fact and the inconsistency of the Al--Ti bonds, it is further speculated 
that the Al--Ti bonds serve to compensate other energy minimization factors for the crystal,
especially those related to symmetry and lattice constraints.

\subsection{Bond expansions} \label{sec:bond-exp}

In Ti$_2$AlC and Ti$_3$AlC$_2$, 
the Al--Ti bonds show the highest expansion [see Table \ref{table:523exp} and Figs. 
\ref{fig:523exp}(b) and \ref{fig:523exp}(c)]. To our knowledge, there are no previous reports of 
temperature-dependent bond lengths in any of the Ti--Al--C MAX phases to which to compare our 
results.  However, in a previous HTND study of the nitride Ti$_2$AlN, the Al--Ti bond also 
showed a higher expansion rate than the Ti--N bond \cite{211paper}. Similarly, a higher expansion 
rate was observed for the $A$--$M$ bonds than the $M$--C bonds in Ti$_3$SiC$_2$ 
\cite{312paper} and Cr$_2$GeC \cite{211paper}.  This result is also consistent with a high-pressure XRD 
study of Ti$_3$AlC$_2$, where the Al--Ti$_\mathrm{I}$ bond was the most compressible, 
while the Ti$_\mathrm{I}$--C and Ti$_\mathrm{II}$--C bonds were more rigid \cite{Volker312}.

In Ti$_5$AlC$_2$, the expansion rate of the Al--Ti$_\mathrm{I}$ bond -- 
14.6 \ex -- is the highest 
of all the bonds in the sample, but the Al--Ti$_\mathrm{III}$ bond expansion is unexpectedly 
low, at 5.5 \ex~ (Table \ref{table:523bonds}).  Nonetheless, the average bond expansion in 523 (10.5 \ex) is still similar  to the 211 and 312 average (11.3 \ex).  Also note that the error bars for the Al--Ti bond expansions in Ti$_5$Al$_2$C$_3$ are the highest of those for the bonds in all phases (see Fig. \ref{fig:523exp}).  This uncertainty further suggests that the Al--Ti bond behavior is flexible within the structure and indicates other crystal imperfections and/or symmetry and lattice dimension effects, as discussed above.

\subsection{Atomic displacement parameters}

The results in Fig. \ref{fig:523adp} show that, like all other MAX phases studied to date, the $A$ atom -- Al in this case -- is a rattler in that it vibrates with a significantly higher amplitude than the other atoms in the structures.  The high ADPs of Al, both calculated and experimental,  relative to the Ti and C atomic displacement values [Figs. \ref{fig:523adp}(a)-(c)] are also consistent with the relatively weaker Al bonding evidenced by the higher Al--Ti expansion rates, at least in Ti$_2$AlC and Ti$_3$AlC$_2$ (Table \ref{table:523bonds}) and the flexibility of the Al--Ti interaction, as discussed above. 

\begin{figure}
\centering
		\includegraphics[width= \columnwidth]{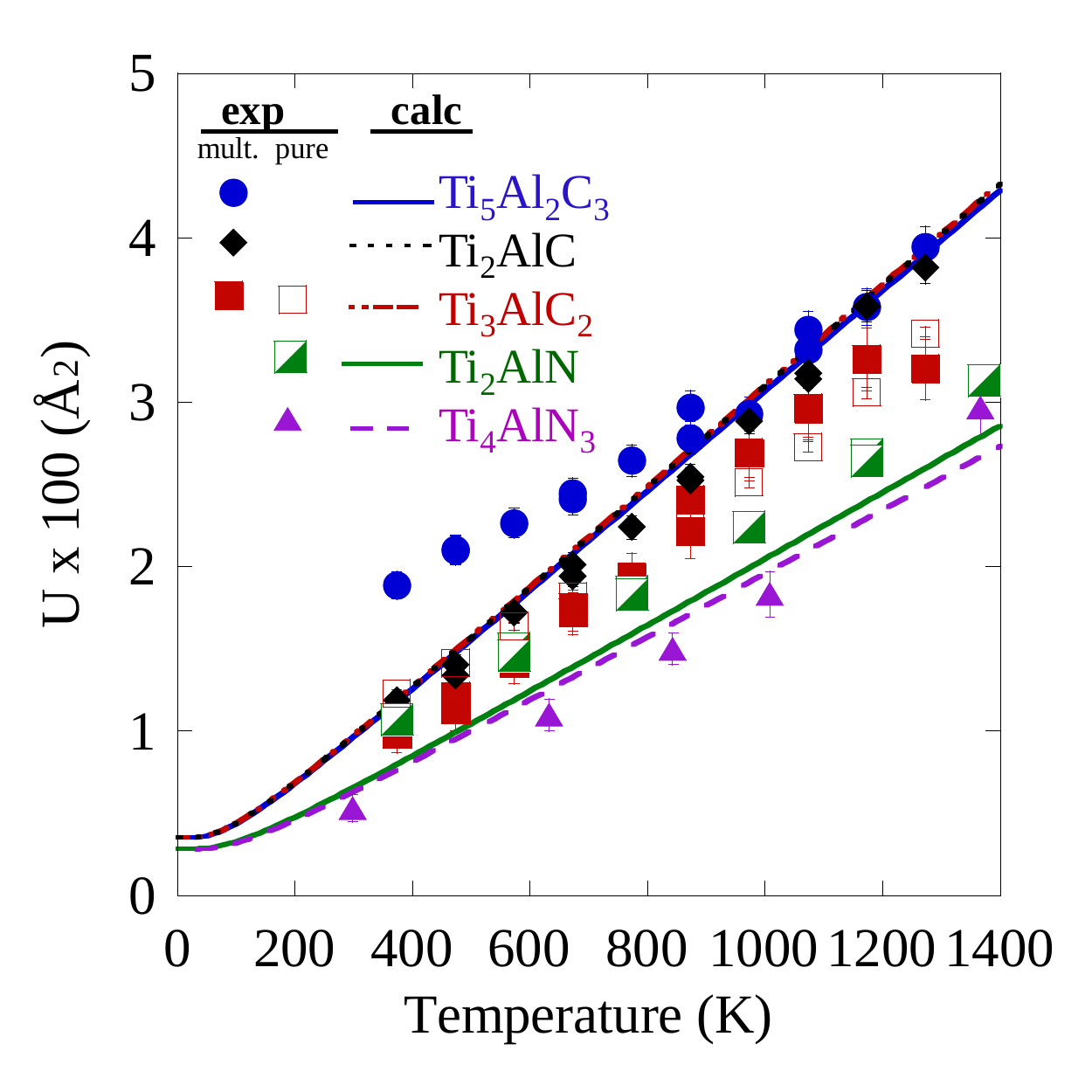}
\caption{
Temperature dependence of isotropic thermal motion $U_\mathrm{iso}$ of Al 
in Ti$_5$Al$_2$C$_3$ (blue circles), Ti$_2$AlC (black diamonds), and Ti$_3$AlC$_2$ (red solid squares) in the multiphase (mult.) sample, and in pure in Ti$_3$AlC$_2$ 
(red open squares) determined from HTND. $U_\mathrm{eq}$ of Al in predominantly single-phase Ti$_2$AlN (green half-filled squares) and Ti$_4$AlN$_3$ (purple triangles) are also shown, determined from HTND in Refs. 
\cite{211paper} and \cite{ThermalTi4AlN3}, respectively.  The lines refer to values calculated
from first-principles phonon calculations, where the calculated values for Ti$_2$AlN and Ti$_4$AlN$_3$ are from Ref. \cite{adpmax}.  Errors are typically smaller than symbol size.
}
 \label{fig:U-Al-all}
\end{figure}

Previous HTND studies of Ti$_3$SiC$_2$ \cite{312paper,Claudia1999}, Ti$_3$GeC$_2$ \cite{312paper}, Ti$_2$AlN \cite{211paper}, Cr$_2$GeC \cite{211paper}, and Ti$_4$AlN$_3$ \cite{ThermalTi4AlN3}  have shown the same rattling phenomenon for the A-group element. A comparison of the vibrational behavior of Al with two other HTND studies of the Al-containing nitrides Ti$_2$AlN and Ti$_4$AlN$_3$ (Fig. \ref{fig:U-Al-all}) further suggests that this ``rattling'' effect is independent of stacking sequence.

The results shown in Figs. \ref{fig:523adp} and \ref{fig:U-Al-all} also clearly indicate that from a theoretical point of view, the ADPs of the three MAX phases should be very comparable.  Given that the  Al atoms in Ti--Al--N nitrides are also predicted to behave similarly to one another -- but different from the carbides -- in their vibrational amplitudes (see lines for Ti$_2$AlN and Ti$_4$AlN$_3$ in Fig. \ref{fig:U-Al-all}), the DFT calculations indicate that the ADPs of Al should not be greatly influenced by stoichiometry.  Interestingly, the agreement between theoretical and experimental isotropic ADPs for the Al atoms is quite good in all five compounds plotted in \ref{fig:U-Al-all}.  The agreement for the other atoms is less good for reasons that are not well understood, but are typical of the MAX phases \cite{adpmax}. 

\section{Summary and Conclusions}

Rietveld analysis of HTND data for a bulk sample containing Ti$_5$Al$_2$C$_3$ [38($\pm$1) wt.\%], Ti$_2$AlC [32($\pm$1) wt. \%], Ti$_3$AlC$_2$ [18($\pm$1) wt. \%], and  (Ti$_{0.5}$Al$_{0.5}$)Al [$12(\pm1)$ wt.\%] 
 has shown that Ti$_5$Al$_2$C$_3$ exhibits similar thermal expansion and thermal motion parameters as Ti$_2$AlC and Ti$_3$AlC$_2$. The thermal expansions for Ti$_5$Al$_2$C$_3$ in the $a$- and $c$- directions, respectively, are $\alpha_a$ = 9.1($\pm 0.1$) \ex and $\alpha_c$ = 9.8($\pm 0.1$) \ex. In all three phases, the average expansion rates of all the Al--Ti bonds are higher than the average Ti--C bond expansions. Ti$_5$Al$_2$C$_3$ consists of alternating layers of 312- and 211-like stacking, where the 312 layers are similar to Ti$_3$AlC$_2$ and the 211 layers are similar to Ti$_2$AlC in dimensions and bond expansions. The Al atoms in all three phases vibrate with higher amplitudes than the Ti and C atoms. This work shows that Ti$_5$Al$_2$C$_3$ exhibits similar properties to Ti$_3$AlC$_2$ and Ti$_2$AlC, two of the most promising MAX phases, which indicates that phase purity can be more relaxed in processing when considering applications. In addition, this work shows that further studies on Ti$_5$Al$_2$C$_3$ can lead to enhanced property optimization and engineering for ternary carbides in the Ti--Al--C system.

\begin{acknowledgments}
This work has benefited from the use of the Lujan Neutron Scattering Center at LANSCE, which is funded by the U.S. Department of Energy's Office of Basic Energy Sciences. Los Alamos National Laboratory is operated by Los Alamos National Security LLC under DOE contract DE-AC52-06NA25396.  This work was also supported by the Army Research Office (W911NF-07-1-
0628).  The authors would also like to thank D. J. Tallman and Dr. T. El-Raghy for providing the samples used in this work. 
\end{acknowledgments}

%


\end{document}